# Passive radiative cooling using temperature-dependent emissivity can sometimes outperform static emitters

*Yeonghoon Jin, Jin-Woo Cho, and Mikhail A. Kats\**

Department of Electrical and Computer Engineering, University of Wisconsin-Madison, Madison, WI, USA

E-mail: mkats@wisc.edu

## Abstract

In passive sky-facing radiative cooling, wavelength-selective thermal emitters in the atmospheric transparency window of 8-13 µm can reach lower temperatures compared to broadband emitters, but broadband emitters always have higher cooling power when the emitter is warmer than the ambient. Here, we propose a temperature-tunable thermal emitter that switches between a wavelength-selective state—with high emissivity only in the atmospheric transparency window of 8-13 µm—and a broadband-emissive state with high emissivity in the 3-25 µm range, thus maintaining high cooling potential across all temperatures. We also propose a realization of such a temperature-tunable emitter using the phase transition of vanadium dioxide ($VO_2$), which can be tuned to the ambient temperature using a combination of doping and defect engineering.

Passive radiative cooling is a technique that lowers an object's temperature without any external energy input, by emitting mid-infrared thermal radiation toward cold outer space.[1,2] The atmosphere has a transparency window in the 8–13 μm wavelength range (**Figure 1a**), and thus the thermal radiation in this window is generally transferred to the cold outer space. At other wavelengths, the atmosphere is only partially transparent due to the absorption of greenhouse gases such as $H_2O$, $CO_2$ and $O_3$, so an object on the surface of the Earth and radiating toward the sky exchanges energy with the moderately cold atmosphere rather than the very cold outer space.[3]

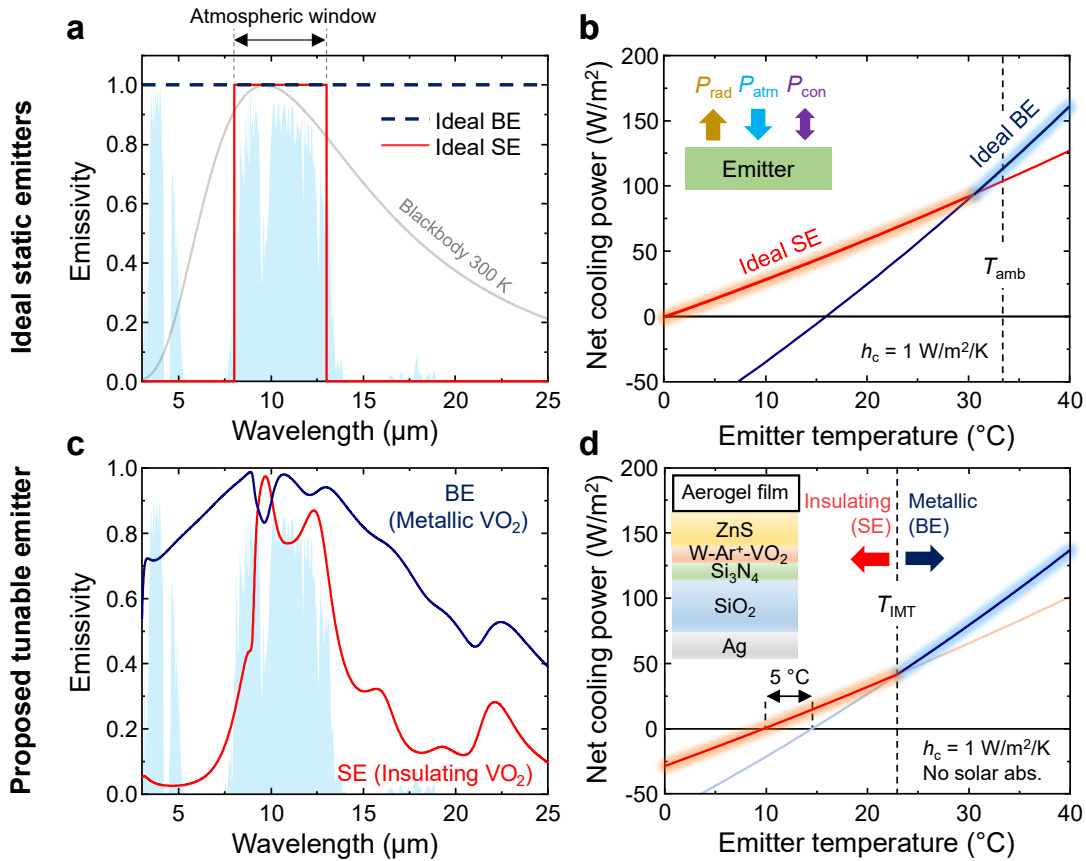

**Figure 1.** Concept of our tunable emitters. **(a)** Emissivity of the ideal wavelength-selective emitter (SE) and the ideal broadband emitter (BE). The shaded region represents the atmospheric transmittance at Cairo on May 1st, 2023 at noon; refer to a paper.[3] The gray line represents the normalized blackbody spectrum at 300 K. **(b)** Net cooling power of the ideal emitters depending on the emitter temperature, assuming no solar absorption and $h_c = 1$ W/m²/K. This graph shows the concept of our tunable emitter, which acts as an ideal SE for the sub-ambient temperature range (<$T_{amb}$), and as an ideal BE for the above-ambient range. **(c)** Emissivity of our proposed tunable emitters for both the insulating (SE) and metallic phases (BE) of tungsten-doped and Ar⁺-irradiated $VO_2$ (W-Ar⁺-$VO_2$). The structure is shown in the inset in Fig. 1d (ZnS 365 nm/W-Ar⁺-$VO_2$ 35 nm/$Si_3N_4$ 100 nm/$SiO_2$ 830 nm/Ag). The aerogel film on top of the structure is employed to minimize convection, assuming 100% transparency in the mid-infrared. **(d)** Net cooling power of the tunable emitter for both phases, assuming the insulator-to-metal transition temperature ($T_{IMT}$) of W-Ar⁺-$VO_2$ is designed to be ~23 °C.

Depending on atmospheric conditions, the temperature of the emitter, and other factors, the ideal thermal emitter may be a wavelength-selective emitter (SE) that has high emissivity in the atmospheric window (8–13 μm) and low emissivity elsewhere, or a broadband emitter (BE) that has high emissivity across a much broader mid-infrared range (e.g., 3–25 μm), as shown in **Figure 1a**. While it is known that the ideal SE is preferred to the ideal BE for sub-ambient cooling,[4] we recently found that, in most practical situations, BEs either match or even outperform SEs.[3]

Nevertheless, one can find combinations of parameters (low humidity, low convective coefficient, and almost no solar absorption) where, at sub-ambient temperatures, SEs have greater cooling power, enabling them to reach meaningfully lower equilibrium temperatures compared to BEs.[5] However, when these same ideal SEs are at or above the ambient temperature ($T_{amb}$), they have lower cooling powers than BEs (**Figure 1b**).

Here, we propose an emitter with temperature-dependent emissivity, switching from selective (SE) at sub-ambient temperatures to broadband (BE) at higher temperatures. This temperature-dependent emitter provides the most cooling power no matter its temperature. For example, **Figure 1b** shows the net cooling power of the ideal SE and BE as a function of emitter temperature, assuming no solar absorption and minimal conduction and convection (non-radiative heat transfer coefficient, $h_c$ = 1 W/m$^2$/K), at noon in Cairo on May 1$^{st}$, 2023; refer to a recent paper[3] for the net cooling power calculation method in a specific location and time. The ideal tunable emitter switches from an ideal SE to an ideal BE at a temperature just below the ambient temperature, $T_{amb}$.

Implementing an emitter with temperature-tunable emissivity requires components with temperature-dependent optical properties. Here, we propose an emitter design based on recently demonstrated tungsten-doped and Ar$^+$-irradiated vanadium dioxide (W-Ar$^+$-VO$_2$).[6] Intrinsic VO$_2$ is a phase-change material that is semiconducting in its low-temperature ("insulating") state, and metallic in its high-temperature state. While its insulator-to-metal transition temperature ($T_{IMT}$) is typically ~70 °C, W doping[7,8] and/or ion irradiation[9,10] can reduce the $T_{IMT}$ to near room temperature.

Our proposed layered temperature-dependent thermal emitter, shown in the inset of **Figure 1d**, comprises of a silver (Ag) substrate, coated with layers of silica (SiO$_2$, 830 nm), silicon nitride (Si$_3$N$_4$, 100 nm), W-Ar$^+$-VO$_2$ (35 nm), and zinc sulfide (ZnS, 365 nm), with an infrared-transparent aerogel film on top. The bottom three layers (Si$_3$N$_4$/SiO$_2$/Ag) form an SE due to the vibrational resonances of SiO$_2$ and Si$_3$N$_4$ in the 8–13 μm range. When the W-Ar$^+$-VO$_2$ is in its insulating state, the entire emitter acts as an SE because of the relatively low loss and small thickness (35 nm) of the W-Ar$^+$-VO$_2$ layer compared to the operating wavelengths. When W-Ar$^+$-VO$_2$ is in its metallic state, it becomes lossy across the mid-infrared and the entire emitter becomes broadband. The ZnS layer acts as an anti-reflection coating to maximize emissivity for both states (refer to **Figure S1** for the refractive indices of the materials used in our structure and **Figure S2** for the emissivity of each case). The aerogel film on top is employed to minimize convection, and we assumed that it can be made transparent in the mid-infrared. The resulting emissivity in both states is shown in **Figure 1c**, and their angle-dependent emissivity spectra are shown in **Figure S3**.

The net cooling power of the designed tunable emitter is shown in **Figure 1d**, assuming no solar absorption and $h_c$ = 1 W/m$^2$/K (refer to a recent paper[3] for details in the cooling power calculation). The ideal $T_{IMT}$ is the crossing temperature between the two emitters, and the tunable emitter acts as an SE below $T_{IMT}$ and an BE above $T_{IMT}$. Although our tunable emitter exhibits some discrepancy from the ideal static emitters in both the emissivities (**Figure 1c** vs. **Figure 1a**) and the cooling powers (**Figure 1d** vs. **Figure 1b**), it still outperforms both the static SE and BE cases with fixed emissivities based on the same structure (the light lines in **Figure 1d**); for example, the equilibrium temperature of the tunable emitter is up to 5 °C lower than that of the static BE case with the same structure.

It is worth noting that this tunable-emitter concept is only effective under conditions of low convection (small $h_c$) and low solar absorption. **Figure 2a** shows the net cooling power of the ideal SE and BE with increasing $h_c$. If no extra effort is taken to suppress convection, a reasonable range for $h_c$ is 6–11 W/m$^2$/K.[11] The superiority of the ideal SE in the sub-$T_{amb}$ range becomes weaker as $h_c$ increases. **Figure 2b** shows the net cooling power of the ideal emitters with solar absorption of 0 and 100 W/m$^2$ at $h_c$ = 1 W/m$^2$/K. The tunable emitter concept does not work if the solar absorption is higher than ~100 W/m$^2$ (roughly ~10% of solar absorption). We previously made these observations when we argued that, in most practical cases, there is no advantage to using SEs for sky-facing radiative cooling.[3] If $h_c$ is higher than 6 W/m$^2$/K and/or the solar absorption is higher than 50 W/m$^2$, the BE state (with metallic W-Ar$^+$-VO$_2$) is always better (**Figure 2c,d**). Therefore, to make use of the advantage of the tunable emitter, both $h_c$ and solar absorption should be as low as possible.

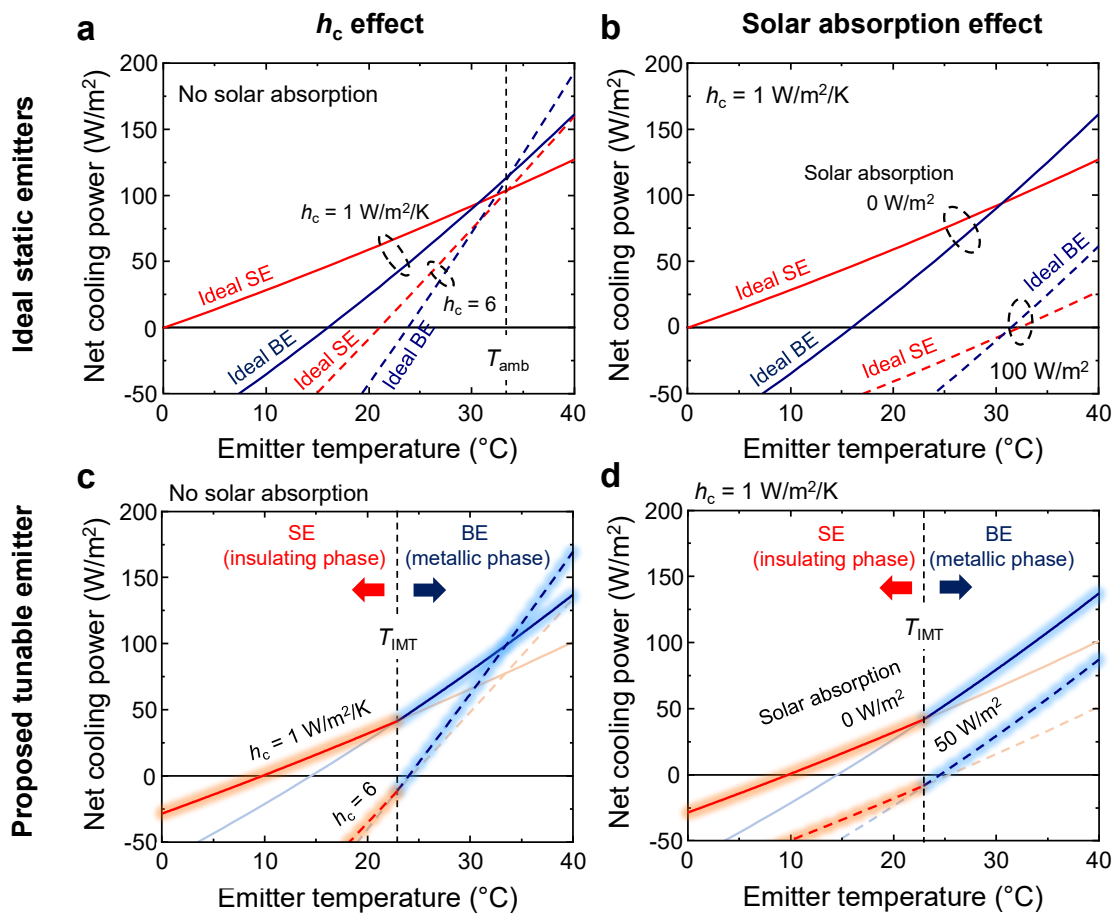

**Figure 2.** The effects of $h_c$ and solar absorption on the cooling performance of thermal emitters. **(a, b)** Net cooling power of the ideal static emitters with **(a)** different $h_c$ (1 and 6 W/m$^2$/K) at no solar absorption, and **(b)** with different solar absorption values (0 and 100 W/m$^2$) at $h_c$ = 1 W/m$^2$/K. **(c, d)** Net cooling power of our tunable emitter with **(c)** different $h_c$ (1 and 6 W/m$^2$/K) at no solar absorption, and with **(d)** different solar absorption values (0 and 50 W/m$^2$) at $h_c$ = 1 W/m$^2$/K. The light, non-highlighted lines represent the net cooling powers of our emitter, assuming VO$_2$ has temperature-independent static emissivity values (either an insulating or metallic phase). The tunable emitter only has an advantage over a static emitter when $h_c$ and solar absorption are both low.

In practice, achieving low $h_c$ (*e.g.*, 1 W/m$^2$/K) and low solar absorption (*e.g.*, <10 W/m$^2$) simultaneously is challenging. Recently, aerogel films that can be attached on top of an emitter have been demonstrated, which exhibit high solar reflection, low convection, and high mid-infrared transparency, though there is a tradeoff between low convection and high infrared transparency,[12] so some additional development is necessary to reach $h_c$ down to 1 and very high transparency.

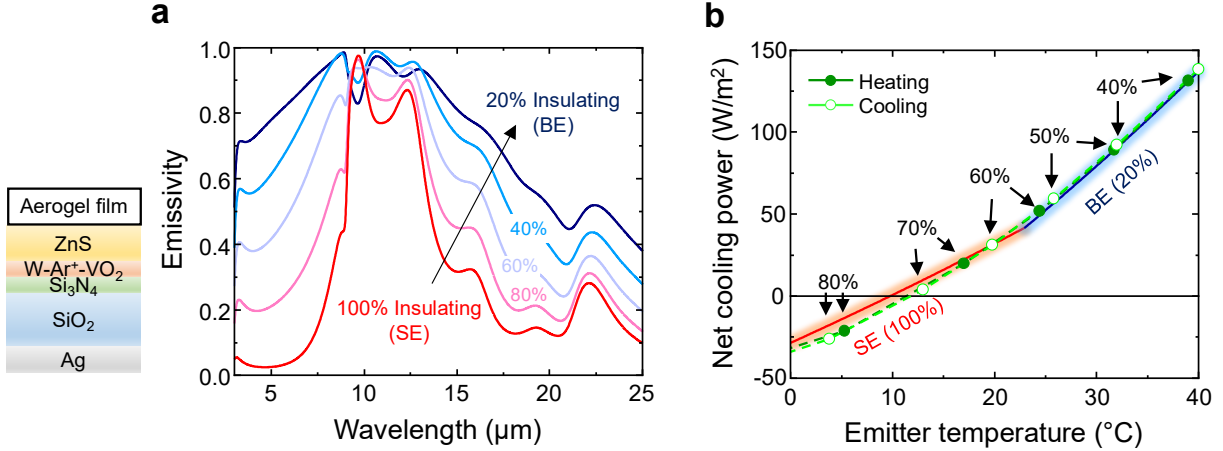

**Figure 3.** The effects of hysteresis and transition temperature width of $VO_2$ on the cooling performance of our tunable emitter. **(a)** Emissivity of our tunable emitter with various mixed phases of $VO_2$. **(b)** Net cooling power of our tunable emitter considering the realistic temperature-dependent optical properties of the W-$Ar^+$-$VO_2$ film shown in Figure S4. We assume no solar absorption and $h_c = 1$ W/m$^2$/K, under conditions in Cario on May 1$^{st}$, 2023, at noon.

One important thing we have not considered so far is the hysteresis of $VO_2$ (between heating and cooling) and the phase-transition temperature width across its phase transition. To account for these, we extracted the temperature-dependent transmittance values of a 1% W-doped and $0.8 \times 10^{14}$ cm$^{-2}$ $Ar^+$-irradiated $VO_2$ film from a recent study[6], for both heating and cooling (**Figure S4**). We also estimated the refractive index of the W-$Ar^+$-$VO_2$ film using an effective medium theory, based on the refractive indices of the insulating and metallic phases of undoped $VO_2$.[13] Note that the refractive index of the low-temperature insulating phase (100% insulating) of the W-$Ar^+$-$VO_2$ film is almost identical to that of undoped $VO_2$, whereas the refractive index of the high-temperature metallic phase of the W-$Ar^+$-$VO_2$ film corresponds to that of a mixed phase (20% insulating and 80% metallic) in undoped $VO_2$. Refer to **Figure S1** for the refractive indices of the W-$Ar^+$-$VO_2$ film with various mixed phases, and **Figure S4** for how we estimated the refractive indices of the W-$Ar^+$-$VO_2$ film with various mixed phases.

Based on the estimated refractive indices of the W-$Ar^+$-$VO_2$ film, we calculated the emissivity of our tunable emitter with mixed phases, from 100% (SE) to 20% (BE) insulating phase (**Figure 3a**). In addition, the net cooling power of the tunable emitter, considering the realistic phase transition of the W-$Ar^+$-$VO_2$ film, is shown in **Figure 3b**; the symbols represent the net cooling power, for both heating and cooling. The solid line represents the net cooling power assuming a (fictitious) abrupt phase transition of the $VO_2$ from the 100% to 20% insulating phase at ~23 °C. Interestingly, the net cooling power curves for heating and cooling overlap. This is because the difference in the percentage of insulating phase between heating and cooling is not significant—only up to 10 %pt of the insulating phase—and this difference has a small impact on emissivity. In addition, the net cooling power of the realistic case almost overlaps with that of the fictitious abrupt phase-transition case (the solid lines). This confirms that the hysteresis and transition-temperature width of $VO_2$ have little effect on the cooling performance of our tunable emitter.

In this paper, we introduced the concept of a temperature-dependent thermal emitter that acts as a selective-emitter (SE) in the sub-ambient temperature range and as a broadband emitter (BE) in the above-ambient range, thus maximizing the radiative cooling potential across all temperature ranges. We proposed a realization of such a tunable emitter based on the phase transition of vanadium dioxide ($VO_2$), where the transition temperature can be selected using a combination of doping and ion irradiation. To fully exploit the advantages of such tunable emitters in practice, a protective film is needed that minimizes convection while maintaining high mid-infrared transparency.

## Acknowledgements

We acknowledge funding from the Office of Naval Research (N00014-20-1-2297) and DARPA (HR00112390123).

## Conflict of Interest

The authors declare no conflict of interest.

## Data Availability

All data used in this work will be uploaded to Zenodo.

# Supporting Information

**Passive radiative cooling using temperature-dependent emissivity can sometimes outperform static emitters**

*Yeonghoon Jin, Jin-Woo Cho, and Mikhail A. Kats\**

Department of Electrical and Computer Engineering, University of Wisconsin-Madison, Madison, WI, USA

E-mail: mkats@wisc.edu

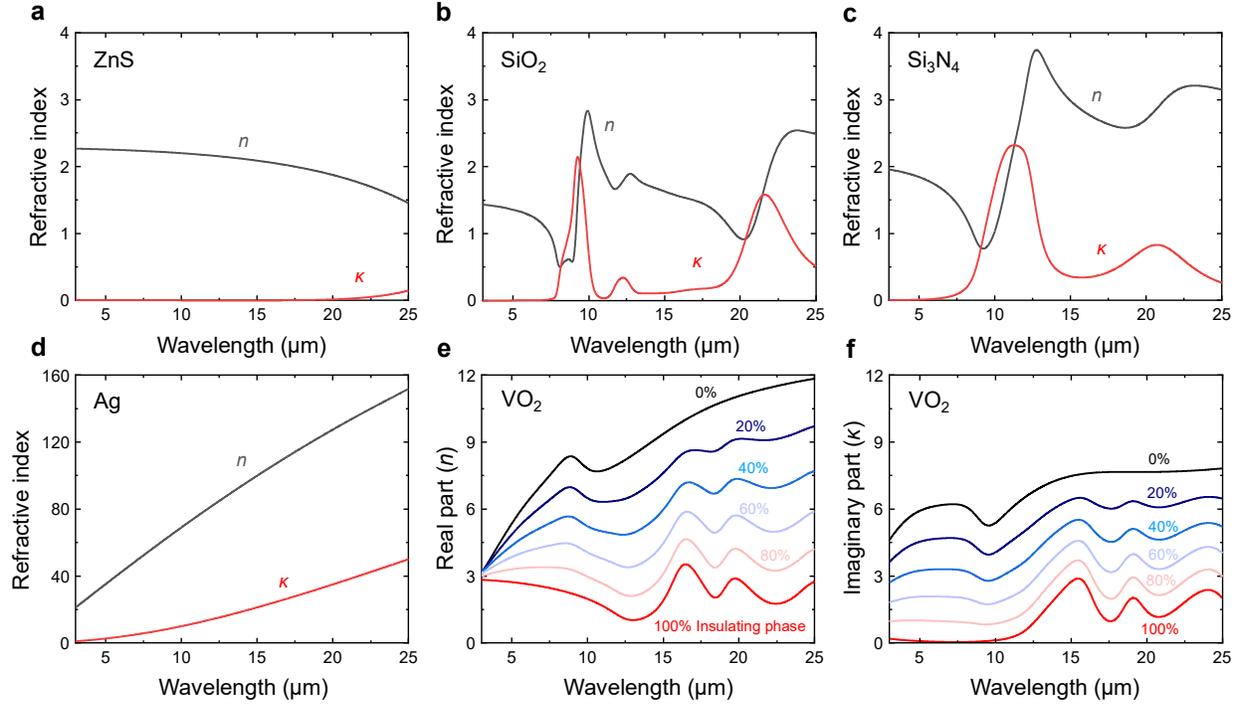

**Figure S1.** Complex refractive indices of the materials used in this work: **(a)** ZnS,[1] **(b)** SiO$_2$,[2] **(c)** Si$_3$N$_4$,[3] and **(d)** Ag.[4] **(e)** The real part (*n*) of the refractive index of VO$_2$ with various portions of the insulating and metallic phases and **(f)** their imaginary part (*κ*). The refractive index of the metallic (the 0% insulating phase) and insulating (100%) phases of undoped VO$_2$ were extracted from the literature.[5] The refractive indices of the mixed phases of VO$_2$ were calculated using an effective medium theory [5,6]; $\tilde{\varepsilon}_{\text{eff}}^{s} = (1-f)\tilde{\varepsilon}_{\text{i}}^{s} + f\tilde{\varepsilon}_{\text{m}}^{s}$, where $f$ is the fraction of the insulating phase, $\tilde{\varepsilon}_{\text{i,m}}$ is the complex permittivity of the insulating and metallic phases, and $s$ varies from -1 to 1 depending on the geometry of metallic inclusions. We used the empirical value of $s = 1$.[6] Note that we used the refractive index of the 100% insulating phase as the refractive index of the low-temperature insulating phase of the W-Ar$^+$-VO$_2$ film, whereas the refractive index of the 20% insulating phase is used to represent the refractive index of the high-temperature metallic phase of the W-Ar$^+$-VO$_2$ film. Refer to Figure S4 regarding the refractive index estimation.

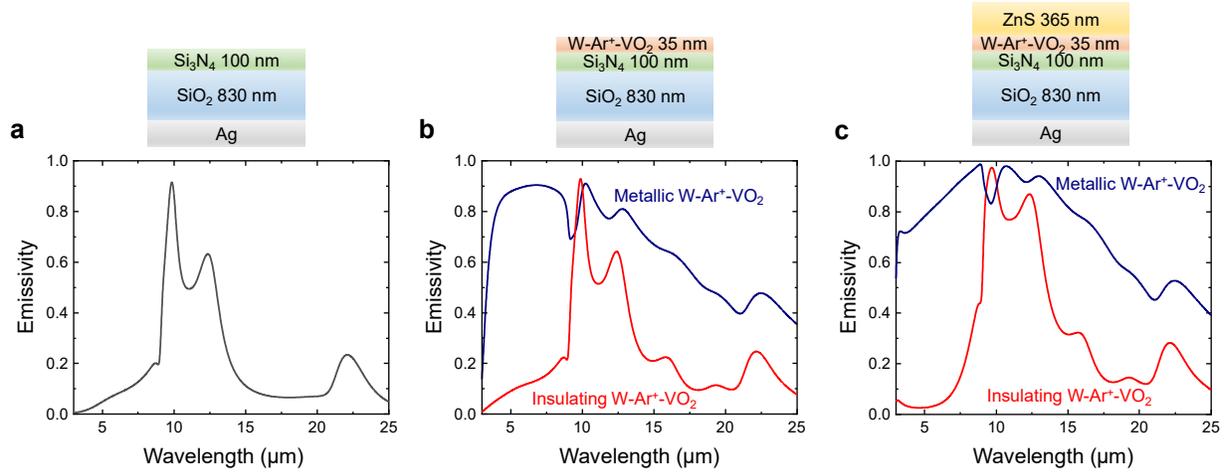

**Figure S2.** Optical design of the tunable emitter. **(a)** Emissivity spectrum of $Si_3N_4$ 100 nm/$SiO_2$ 830 nm/Ag. The wavelength-selective emission at 8–13 μm originates from the vibrational resonance modes of $Si_3N_4$ and $SiO_2$, as shown in **Figure S1b, c**. **(b)** Emissivity spectra of the structure in (a), but with 35 nm of W-$Ar^+$-$VO_2$ on top, with the W-$Ar^+$-$VO_2$ in its metallic and insulating phases. When the W-$Ar^+$-$VO_2$ is in the insulating phase, where the extinction coefficient ($\kappa$) is not significant (**Figure S1f**) and its thickness is very thin compared to mid-infrared wavelengths, the entire emitter maintains selective emission. When the W-$Ar^+$-$VO_2$ is in the metallic phase, where the extinction coefficient is large (**Figure S1f**) across the mid-infrared, the entire emitter becomes broadband. **(c)** Emissivity spectra the structure in (b), but with an additional 365 nm of ZnS on top. The ZnS is mid-infrared transparent (**Figure S1a**), and it acts as an anti-reflection coating.

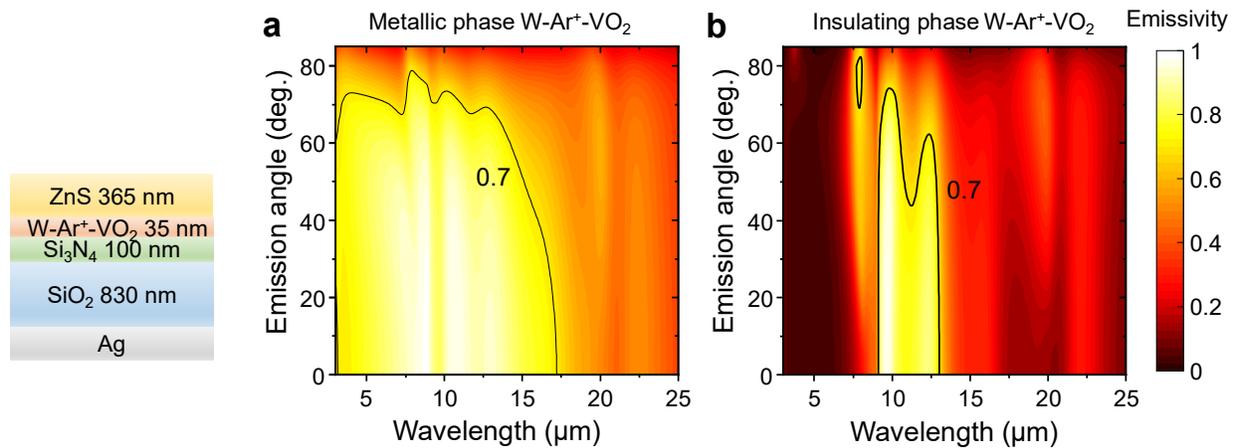

**Figure S3.** Calculated polarization-averaged angle-dependent emissivity of the tunable emitter using the transfer-matrix method, when the W-$Ar^+$-$VO_2$ is in **(a)** the metallic phase and **(b)** the insulating phase. The emitter maintains high emissivity (>70%; see contour in both panels) up to an emission angle of 70 degrees.

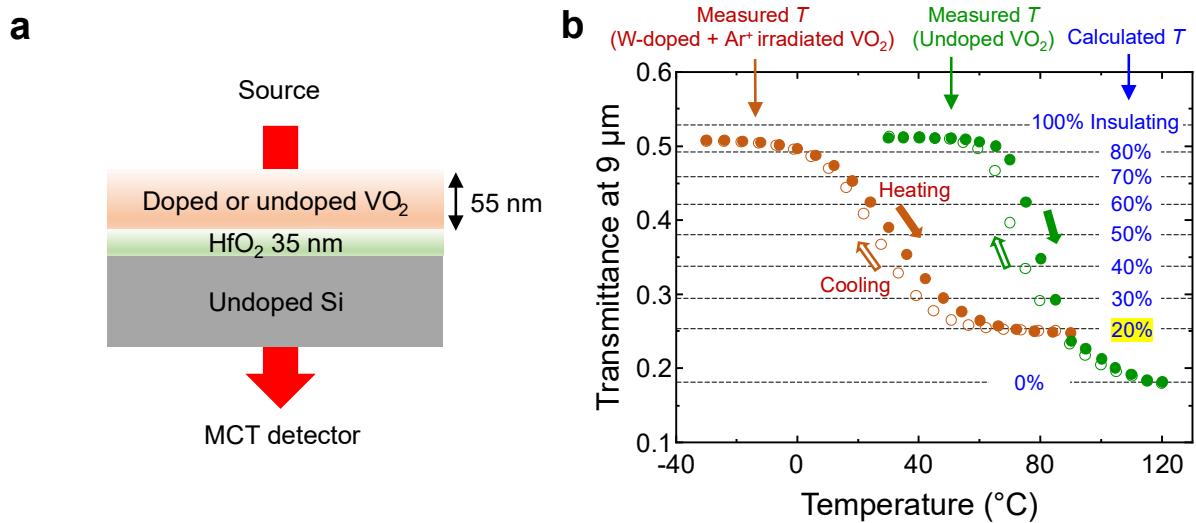

**Figure S4. (a)** Illustration of the transmittance measurement experiment. **(b)** Measured transmittance of an 1% W-doping and 0.8×10$^{14}$ cm$^{-2}$ Ar$^+$ irradiated VO$_2$ film and an undoped VO$_2$ film (on HfO$_2$/Si), for cooling and heating. The experimental data was extracted from a recent paper[7].

We extracted the transmittance values of an 1% W-doped and $0.8 \times 10^{14}$ cm$^{-2}$ Ar$^+$-irradiated VO$_2$ film and an undoped VO$_2$ film from a recent study[7], for both heating and cooling (**Figure S4b**). The corresponding experimental setup and the sample configuration are shown in **Figure S4a**. The transmittance values of the low-temperature insulating phase for both samples are almost the same, so it is reasonable to use the refractive index of undoped VO$_2$ to represent that of the W-Ar$^+$-VO$_2$ in the insulating phase. However, the transmittance values of the high-temperature metallic phase for both samples are noticeably different.

As a result, we estimated the refractive index of the W-Ar$^+$-VO$_2$ film using an effective medium theory, based on the refractive indices of the insulating and metallic phases of undoped VO$_2$ [5]. To explain in detail, we calculated the transmittance values of the structure (VO$_2$ 55 nm/HfO$_2$ 35 nm/Si) using the transfer-matrix method, with various mixed phases of VO$_2$ (**Figure S4b**). The calculated transmittance values of the full insulating and metallic (0% insulating) phases show good agreement with the experimental data of the undoped VO$_2$ sample (0.53 for the insulating phase and 0.18 for the metallic phase). In addition, we found that the 20% insulating phase corresponds to the high-temperature metallic phase of the W-Ar$^+$-VO$_2$ film. As a result, we used the refractive index of the 20% insulating phase (assuming intrinsic VO$_2$) as the refractive index of the metallic phase of the W-Ar$^+$-VO$_2$ film (**Figure S1e,f**). In the same way, we can estimate the refractive index of the W-Ar$^+$-VO$_2$ film with various mixed phases (**Figure S1e,f**).